\documentclass{iau}
\usepackage{graphicx}
\newcommand{\hagn}{\mbox{{\sc \small Horizon-AGN\, }}}

\title[Intrinsic alignments of galaxies] 
{How the cosmic web induces intrinsic alignments of galaxies}

\author[S. Codis et al.]  
{S. Codis$^{1}$\thanks{Email: {\tt codis@iap.fr} },
Y. Dubois$^{1}$,
C. Pichon$^{1}$,
J. Devriendt$^{2}$ 
and A. Slyz$^{2}$
}
\affiliation{
$^{1}$Sorbonne Universit\'es, UPMC Univ. Paris 06 \& CNRS, UMR7095, \\Institut d'Astrophysique de Paris, 98 bis Boulevard Arago, 75014, Paris, France \\[\affilskip]
$^{2}$Sub-department of Astrophysics, University of Oxford, Keble Road, Oxford OX1 3RH
}

\pubyear{2014}
\volume{308}  
\pagerange{119--126}
\jname{The Zeldovich Universe,
Genesis and Growth of the Cosmic Web}
\editors{Rien van de Weygaert, Sergei Shandarin, Enn Saar \& Jaan Einasto}
\begin{document}

\maketitle

\begin{abstract}
Intrinsic alignments are believed to be a major source of systematics for future generation of weak gravitational lensing surveys like Euclid or LSST.
Direct measurements of the alignment of the projected light distribution of galaxies in wide field imaging data seem to agree on a contamination at a level of a few per cent of the shear correlation functions, although the amplitude of the effect depends on the population of galaxies considered. 
Given this dependency, it is difficult to use dark matter-only simulations as the sole resource to predict and control intrinsic alignments. 
We report here estimates on the level of intrinsic alignment in the cosmological hydrodynamical simulation \hagn that could  be a major source of systematic errors in weak gravitational lensing measurements. In particular, assuming that the spin of galaxies is a good proxy for their ellipticity, we show how those spins are spatially correlated and how they couple to the tidal field in which they are embedded. We will also present  theoretical calculations that illustrate and qualitatively explain the observed signals. 
\end{abstract}

\firstsection 
\section{Introduction}

Weak lensing is often presented as a potential powerful probe of cosmology for the coming years with 
 large surveys like DES,
 Euclid 
 or LSST.
It relies on the fact that the observed shape of galaxies is distorted because the light path from background sources towards us is bent by the gravitational potential well along the line of sight. Therefore, measuring these distortions directly probes cosmology (cosmological model, dark matter distribution, etc). The idea behind weak lensing cosmic probes  is thus to try and detect coherent distortions of the shapes of galaxies, e.g using the two-point correlation function of the ellipticities of galaxies. Note that the apparent ellipticity of a galaxy is induced by the cosmic shear $\gamma$ (which is related to the projected gravitational potential along the line of sight) but also encompasses the intrinsic ellipticity of that galaxy
$
  e = e_s + \gamma\;,
$
where $e$ is the apparent ellipticity and $e_s$ the intrinsic source ellipticity (that would have been observed without lensing).
 Therefore the (projected) ellipticity-ellipticity two-point correlation function can be written as the sum of a shear-shear term, intrinsic-intrinsic and intrinsic-shear correlations
 \begin{equation}
\label{eq:ee}
\left\langle e(\vartheta) e(\vartheta+\theta) \right\rangle_\vartheta = \left\langle \gamma \gamma' \right\rangle+\left\langle e_{s}e_{s}' \right\rangle+2\left\langle e_{s} \gamma'\right\rangle
\,,
\end{equation}
where, for compactness, the prime means at an angular distance $\theta$ from the first location. 
These last two contributions that contaminate the shear signal are the two kinds of intrinsic alignments (IA hereafter), one term being the so-called ``II''  term $\left\langle e_{s}e_{s}' \right\rangle$ induced by the intrinsic correlation of the shape of galaxies in the source plane (\cite{HRH00,C+M00,Cat++01}) and the other one is the so-called ``GI'' term $\left\langle e_{s} \gamma'\right\rangle$ coming from correlations between the intrinsic ellipticity of a galaxy and the induced ellipticity (or shear) of a source at higher redshift (\cite{H+S04}).
Much effort has thus been made to control the level of IA of galaxies as a potential source of systematic errors in weak gravitational lensing measurements although some techniques have been proposed to mitigate their nuisance by making extensive use of photometric redshifts (e.g. \cite{Bla++12}).
Direct measurements of the alignment of the projected light distribution of galaxies in wide field imaging data seem to agree on a contamination at a level of a few percents in the shear correlation functions, although the amplitude of the effect depends on the depth of observations, the amount of redshift information and the population of galaxies considered in the sense that red galaxies seem to show a strong intrinsic projected shape alignment signal whereas observations only place upper limits in the amplitude of the signal for blue galaxies (e.g. \cite{Joa++13a}).

\begin{figure}[b]
\begin{center}
\includegraphics[width=2.5in]{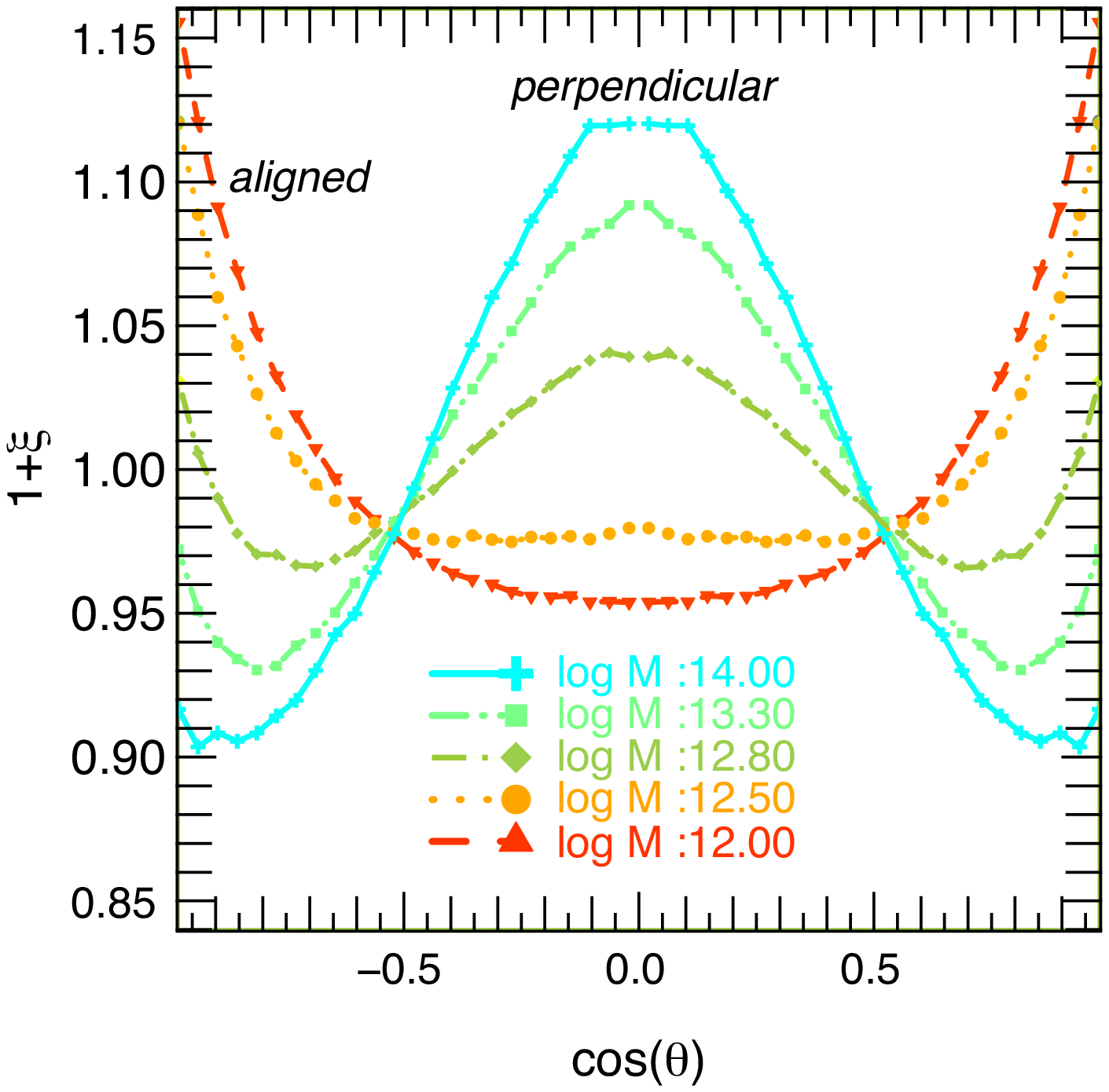} 
\includegraphics[width=2.5in]{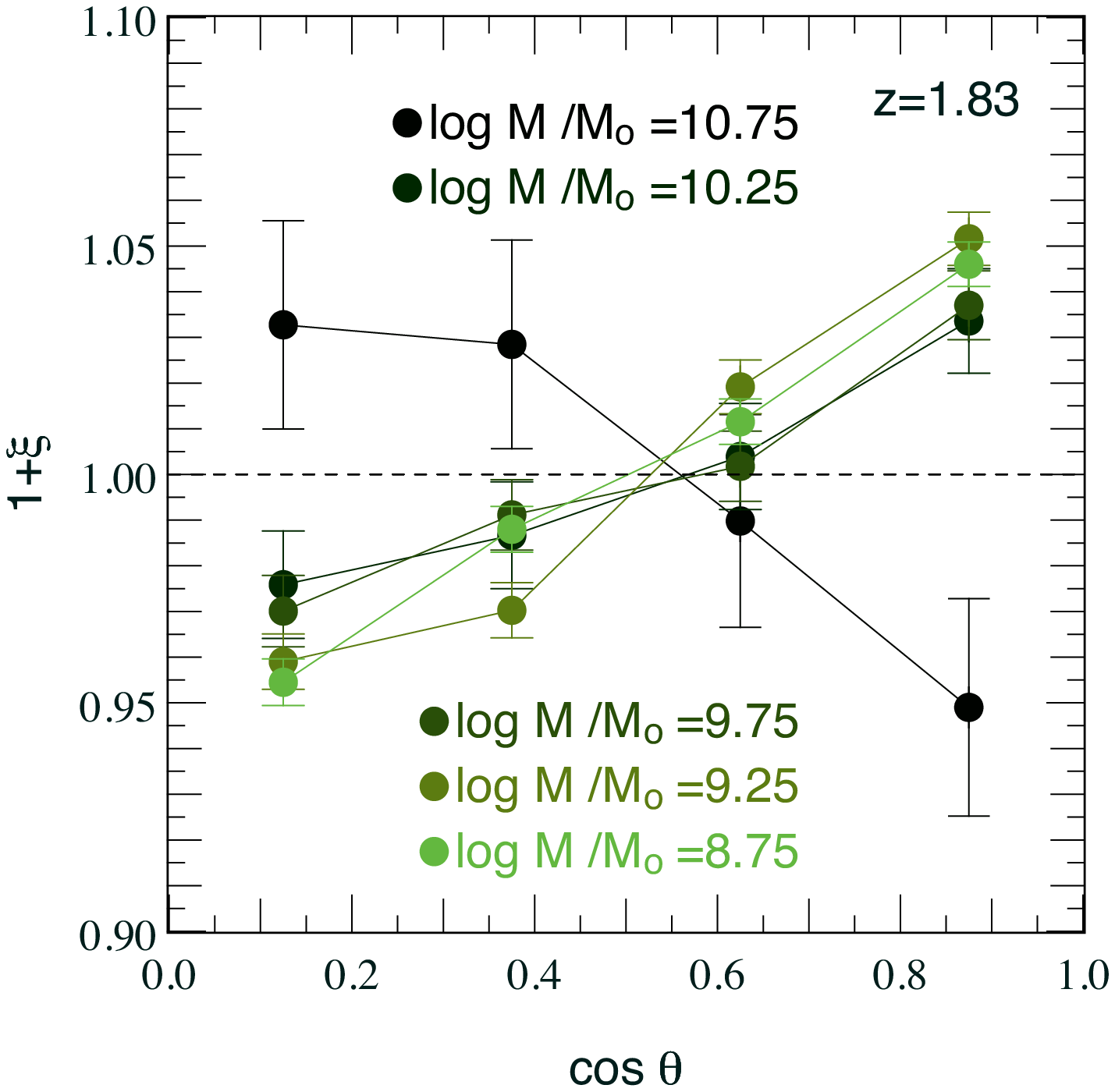} 
\caption{{\sl Left:} 
  excess probability of alignment  between the spin and the direction of the closest filament as measured from the 43 millions haloes of  the Horizon 4$\pi$ simulation (\cite{teyssier02}) at redshift zero. 
  Different colours correspond to different mass bins from $ 10^{12}$({ red}) to $ 10^{14}$ $M_\odot$ ({blue})  as labeled.
   A transition  mass is detected at $M_{0}^{s}= M_{\rm crit}^{s}(z=0)\simeq 5 (\pm 1)\times 10^{12} M_{\odot}$: for haloes with $M>M_{0}^{s}$,  the spin is more likely to be perpendicular to their host filament, whereas for haloes with $M<M_{0}^{s}$, the spin tends to be aligned with the closest filament.  This figure is from \cite{codisetal12}.
    {Right-hand panel:} 
    same as left panel for the 160 000 galaxies of the \hagn hydrodynamical simulation at $z=1.8$ (\cite{Dubois14}).
}
   \label{fig1}
\end{center}
\end{figure}

From a theoretical point of view, it has been shown that dark halos (\cite{calvoetal07, pazetal08,codisetal12} among many others) and galaxies (\cite{Hahn10,Dubois14}) are correlated with the cosmic web. Fig.~\ref{fig1} shows that the spin of dark halos (left panel) and galaxies (right panel) is correlated to the direction of the closest filament. Consequently, this large-scale coherence of galaxies could then contaminate significantly the weak lensing observables.
Given the inherently anisotropic nature of the large-scale structure and its complex imprint on the shapes and spins of galaxies 
together with the dependency on the physical properties of the galaxies seen in the observation, it is probably difficult to rely on isotropic linear theory (e.g. \cite{Lee01}) or dark matter-only numerical simulations as the sole resort to predict and control IA for weak lensing applications. With the advent of cosmological hydrodynamical simulations, we are now in a position to try and measure IA directly into those simulations instead of relying on pure N-body simulations and semi-analytical models (\cite{S+B10,Joa++13b}). We  report on the recent findings of \cite{codisetal14} who uses the \hagn simulation presented in \cite{Dubois14} at redshift $z=1.2$ to measure the level of IA taking the spin as a proxy for the shape of galaxies.
Section~2 will be devoted to the measured correlations between galaxy shapes and tidal field (related to the ``GI'' term) and section~3 to the auto-correlation of the intrinsic ellipticities (related to the ``II'' term).

\section{Gravitational-intrinsic correlations}
In order to study the correlations between the spin direction and the surrounding tidal field, the traceless tidal shear tensor is computed in the \hagn simulation
$
  T_{ij} = \partial_{ij} \Phi -\Delta \Phi \,{\delta}_{ij}/3\,,
$
$\Phi$ being the gravitational potential and $\delta_{ij}$ the Kronecker delta function. 
The minor, intermediate and major eigen-directions of the tidal tensor $T_{ij}$ are called
$\mathbf{e}_1$, $\mathbf{e}_2$ and $\mathbf{e}_3$ corresponding to the ordered eigenvalues 
$\lambda_{1}\le\lambda_{2}\le\lambda_{3}$ 
of the Hessian of the gravitational potential,
$\partial_{ij} \Phi$.
In the filamentary regions, 
$\mathbf{e}_1$ gives the direction of the filament, while the walls
are collapsing along $\mathbf{e}_3$ and extend, locally, in the 
plane spanned by $\mathbf{e}_1$ and $\mathbf{e}_2$ (\cite{Pogosyanetal1998}). 

\subsection{One-point correlation between spin and tidal tensor}
\begin{figure}[b]
\begin{center}
\includegraphics[width=2.5in]{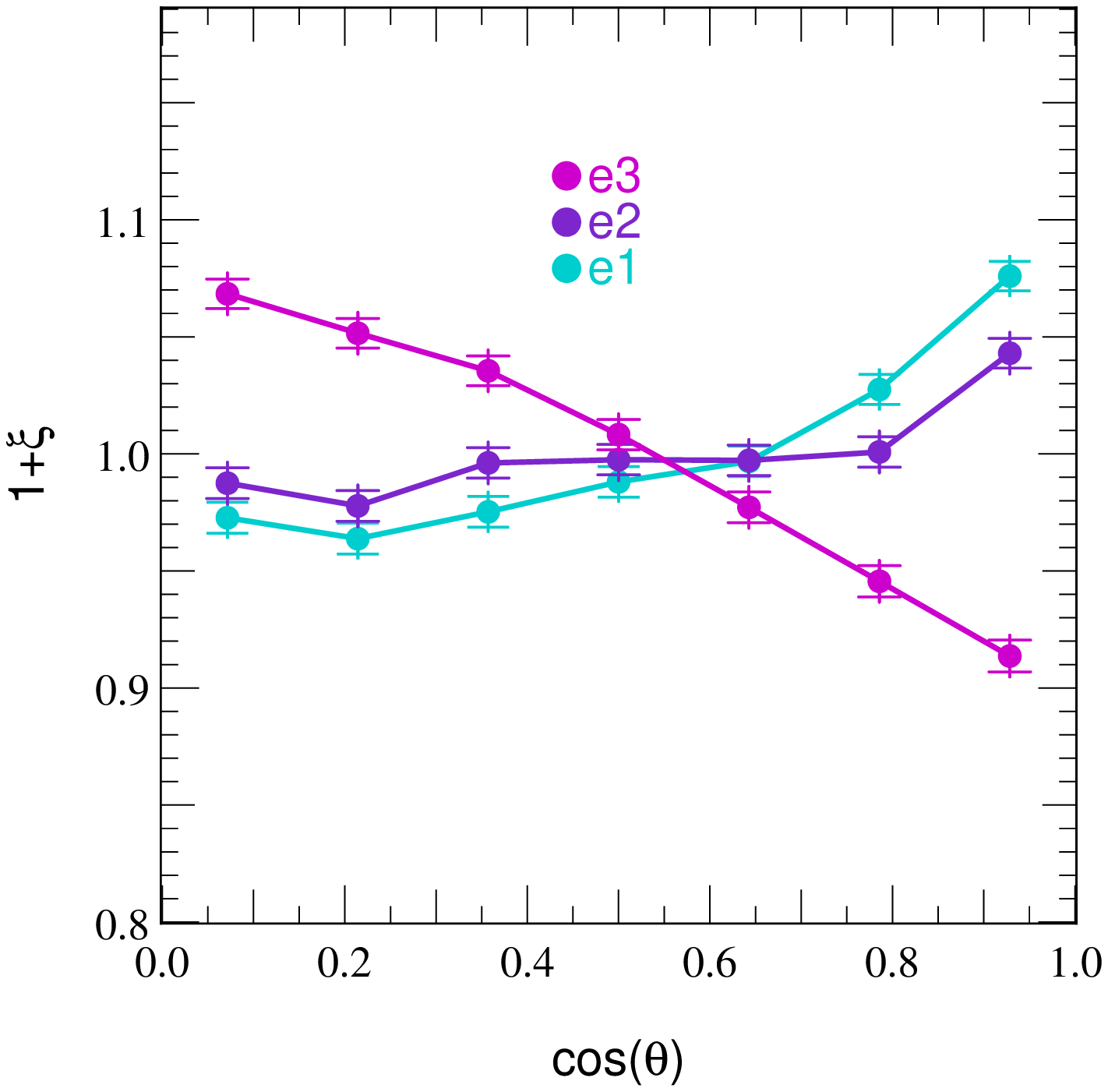}
\includegraphics[width=2.5in]{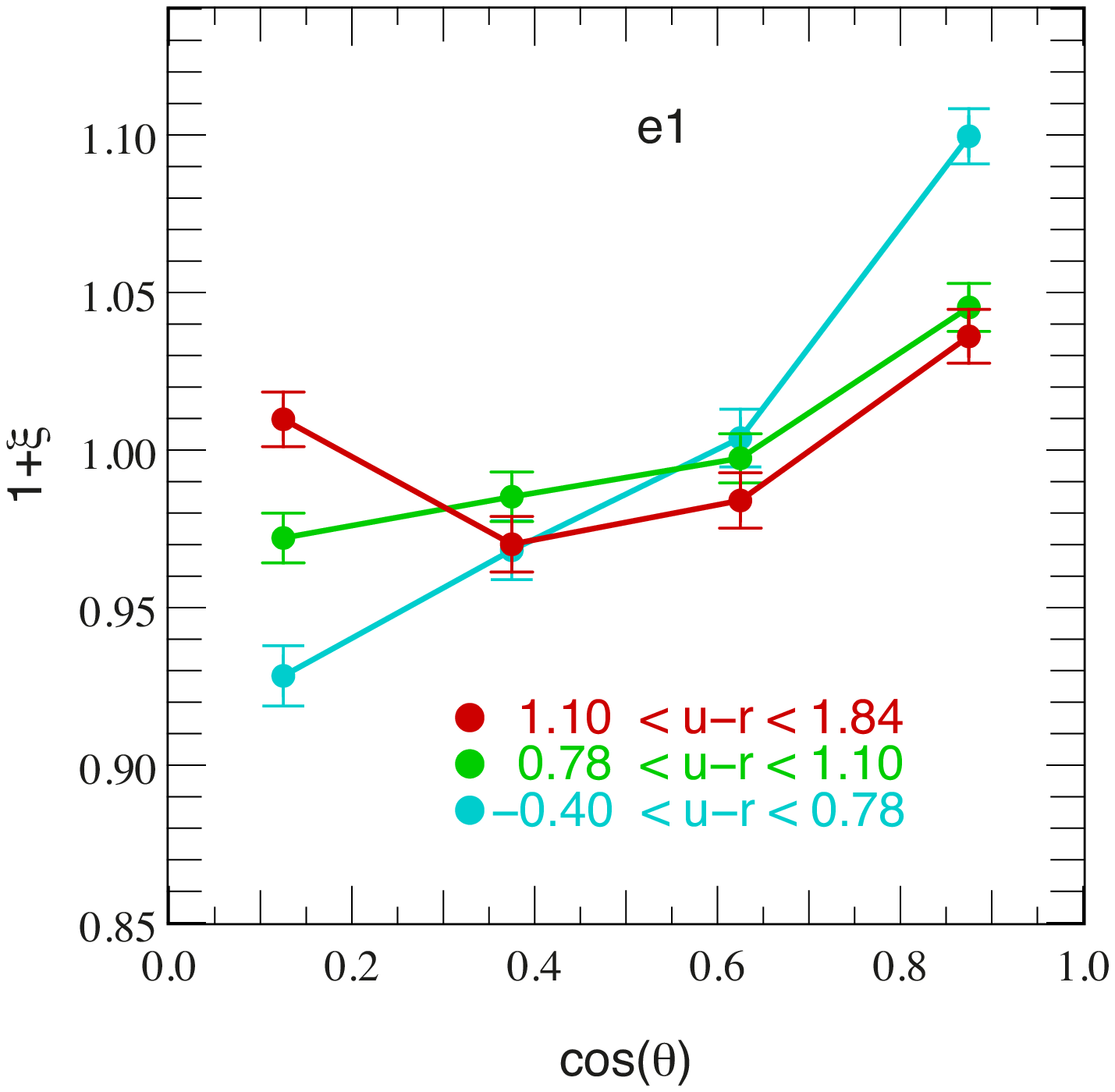}
\caption{
{\sl Left:} PDF of the cosine of the angle between the spin of galaxies and the minor (cyan), intermediate (purple) and major (magenta) eigen-direction of the tidal tensor in the \hagn simulation. {\sl Right:} PDF of the cosine of the angle between the spin of galaxies and the minor eigen-direction for different colours as labeled.
}
   \label{fig2}
\end{center}
\end{figure}
The cosine of the angle between the spin of the galaxies and the three local eigen-directions of the tidal tensor is then measured. An histogram of these values is computed and rescaled so as to give the probability distribution function (PDF) displayed in Fig.~\ref{fig2} (left panel). The spins clearly exhibits a tendency to be aligned with the minor eigen-direction (i.e the filaments) and in a weaker way with the intermediate axis (i.e the wall). The same analysis can be done for different mass and colour samples (see Fig.~\ref{fig2}, right panel) : small-mass galaxies tend to have a spin aligned with the minor eigen-direction while more massive galaxies tend to have their spin perpendicular to that direction. The transition seems to occur around $4\times 10^{10}M_{\odot}$. With regards to colours,  the bluest galaxies (defined here by $u-r<0.78$) are more correlated with the tidal eigen-directions than the red galaxies ($u-r>1.1$). This can be easily understood as red galaxies are typically massive, while blue galaxies are often small-mass galaxies. At that redshift ($z\sim 1.2$), this implies that red galaxies correspond to objects around the transition mass, whereas blue galaxies are mostly aligned with $\mathbf{e}_{1}$. At lower redshift, we expect the population of massive galaxies perpendicular to $\mathbf{e}_{1}$ to increase, so that red galaxies become more correlated. Obviously, we should also keep in mind that applying additional selection cuts on the galaxy samples (mass, luminosity, etc) would change the level of correlation.

\subsection{Two-point correlation between spin and tidal tensor}
\begin{figure}[b]
\begin{center}
\includegraphics[width=2.52in]{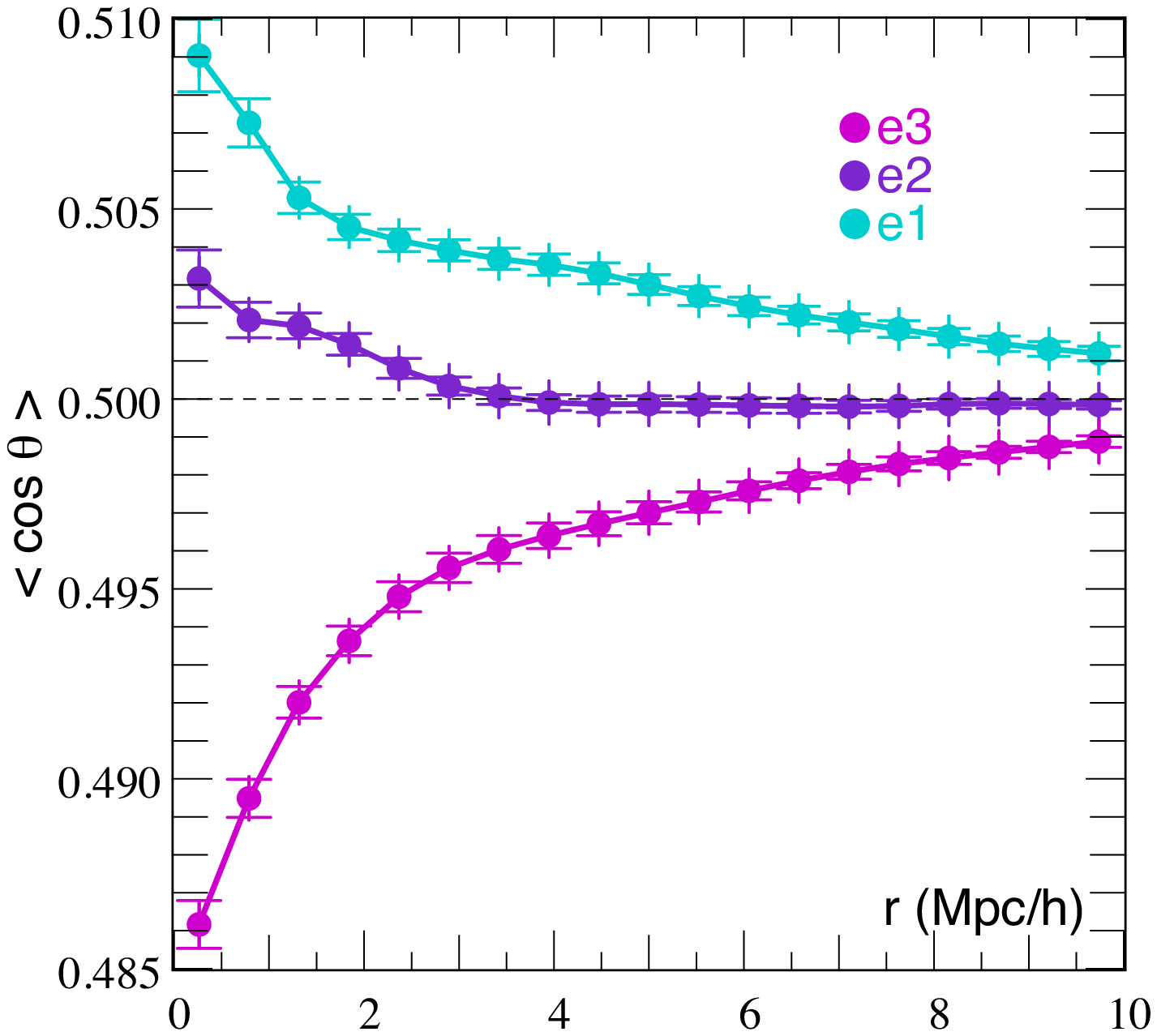}
\includegraphics[width=2.4in]{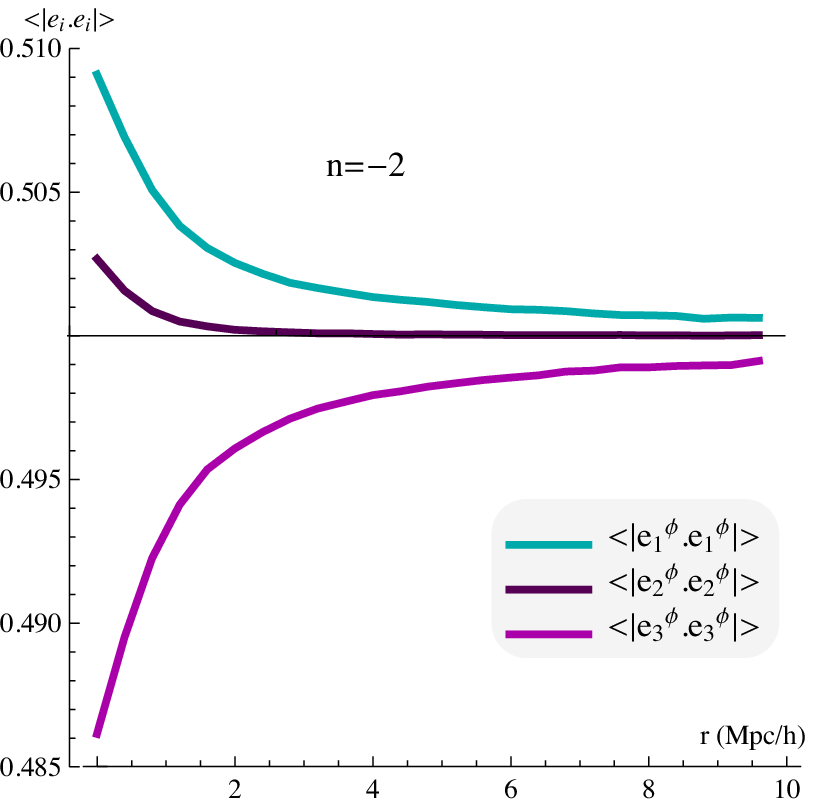}
\caption{
{\sl Left:} Mean cosine of the angle between the spin of galaxies and the minor (cyan), intermediate (purple) and major (magenta) eigen-direction of the tidal tensor as a function of the separation. {\sl Right:} Same as left panel for a Gaussian random field with power-law power spectrum once rescaled so as to match the measured value at zero separation. 
}
   \label{fig3}
\end{center}
\end{figure}
Beyond one-point statistics, it is also of interest in the context of weak lensing studies to quantify how this signal pervades when the tidal field at a distance $r$ from the galaxy is considered.
Since the tidal field in the vicinity of a galaxy contributes also to the lensing signal carried by more distant galaxies, it is  clear that the spin -- tidal tensor cross-correlation is closely related to the GI term. 
In order to o address that question, the correlations between the spins and the eigen-directions of the tidal tensor at comoving distance $r$ are measured. 
 Fig.~\ref{fig3} (left panel) shows the mean angle between the spins and the three eigen-directions of the tidal field $\mathbf{e}_{1}$ (cyan), $\mathbf{e}_{2}$ (purple) and  $\mathbf{e}_{3}$ (magenta) as a function of the separation $r$.
As expected, the spin and the tidal eigen-directions de-correlate with increasing separation. However, whereas the signal vanishes on scales $r > 3$ $h^{-1}\,Ê\rm Mpc$ for the spin to intermediate tidal eigen-direction correlation, it persists on distances as large as $\sim 10$ $h^{-1}\,Ê\rm Mpc$ for the minor and major eigen-directions of the tidal tensor. This behaviour can be theoretically understood (Codis et al, \textit{in prep}) using a Gaussian random field $\delta$ (here we use a power-law power spectrum with spectral index $n=-2$) for which we compute the joint PDF of the second derivatives of its corresponding potential ($\phi_{ij}$, $\phi$ being related to $\delta$ by the Poisson equation). Then the mean angle between the eigen-directions of $\phi_{ij}$  in two locations separated by $r$ can be computed. Once rescaled so as to match the one-point statistics measured in Fig.~\ref{fig2} (here we want to study the evolution of the two-point function with the separation, not its absolute value), we find the function plotted on the right panel of Fig.~\ref{fig3} which interestingly shows the same qualitative behaviour as what is measured in the simulation (left panel).

\section{Intrinsic-intrinsic correlations}
\subsection{3D spin-spin correlations}
\begin{figure}[b]
\begin{center}
\includegraphics[width=2.5in]{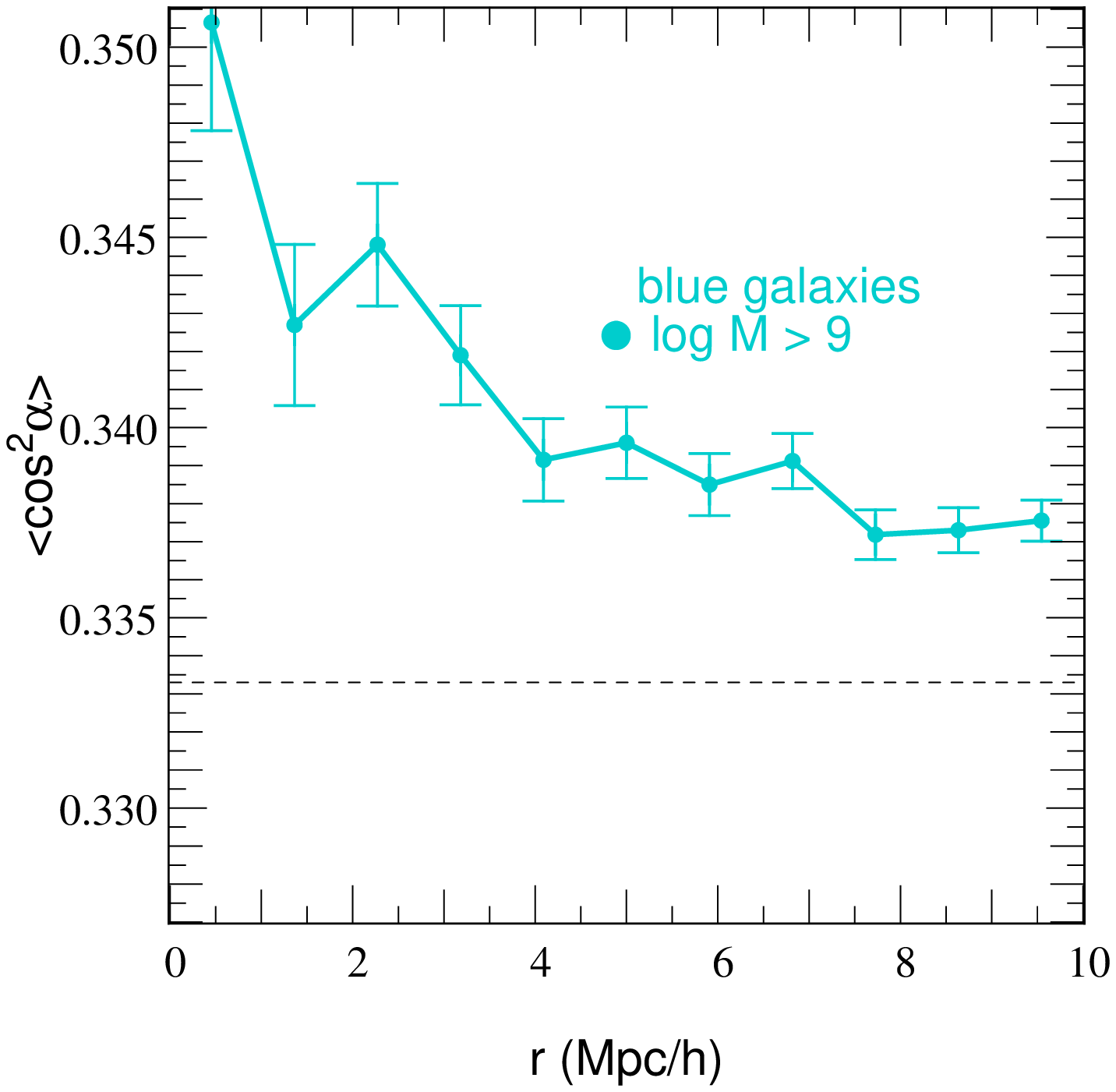}
\includegraphics[width=2.5in]{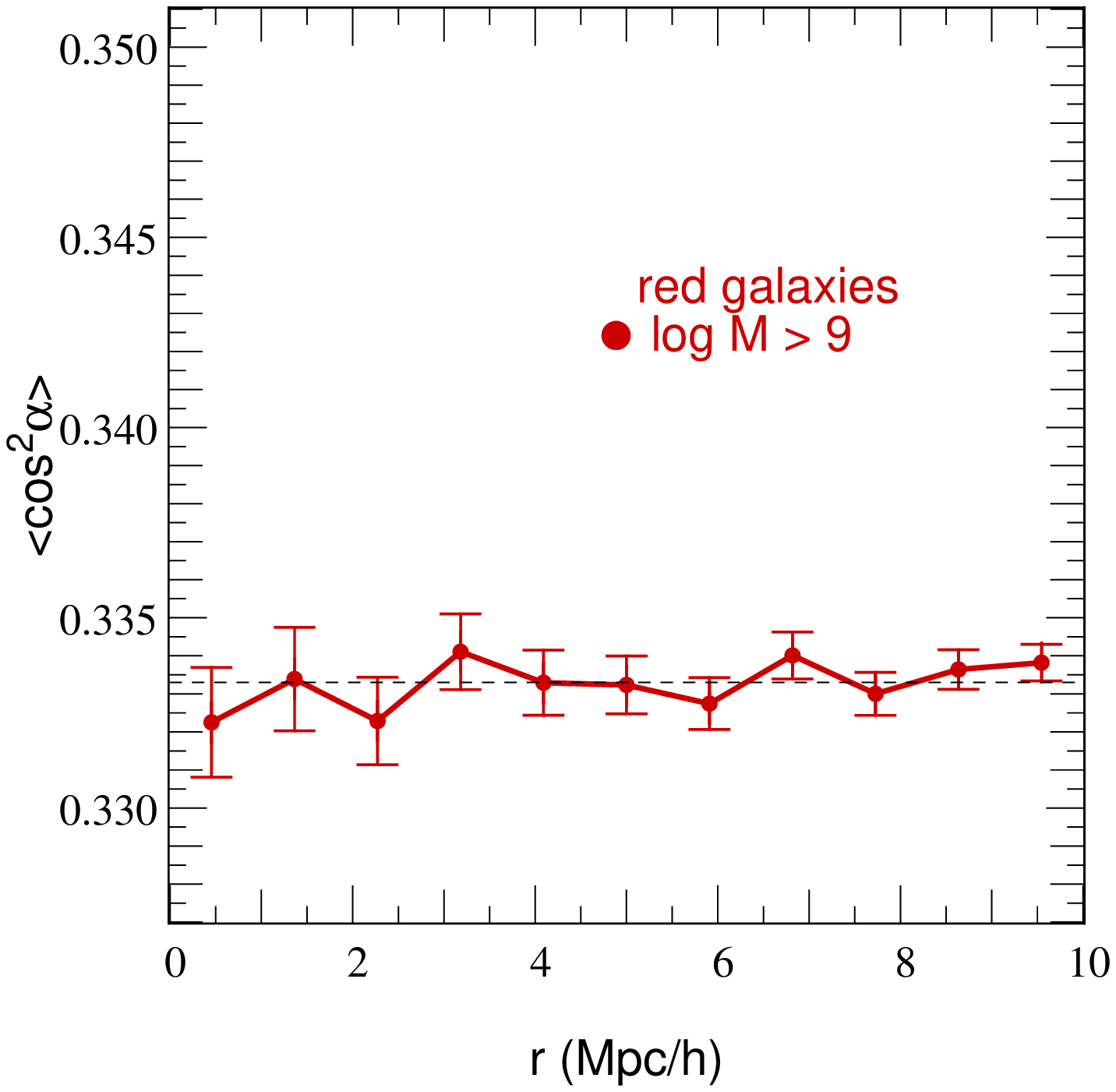}
\caption{
{\sl Left:} mean cosine square of the angle between the spin of blue galaxies separated by $r$. {\sl Right:} same as left panel for red galaxies. Important correlations on cosmological scales are detected for the blue sample, not for red galaxies.
}
   \label{fig4}
\end{center}
\end{figure}
The IA contamination coming from the auto-correlations of the intrinsic ellipticities is now investigated by means of the spin-spin correlation function.
The angle between the spins of each pair of galaxies separated by a distance $r$ is computed and the resulting histogram of the square of the cosine of those angles (the polarity is not relevant for weak lensing studies) is shown in Fig.~\ref{fig4} for two different colour samples. A significant spin correlation is detected for blue galaxies out to at least 10 Mpc/h. Conversely, we detect no significant correlations for red galaxies at that redshift ($z=1.2$). 

\subsection{Projected spin-spin correlations}
\begin{figure}[b]
\begin{center}
\includegraphics[width=1.7in]{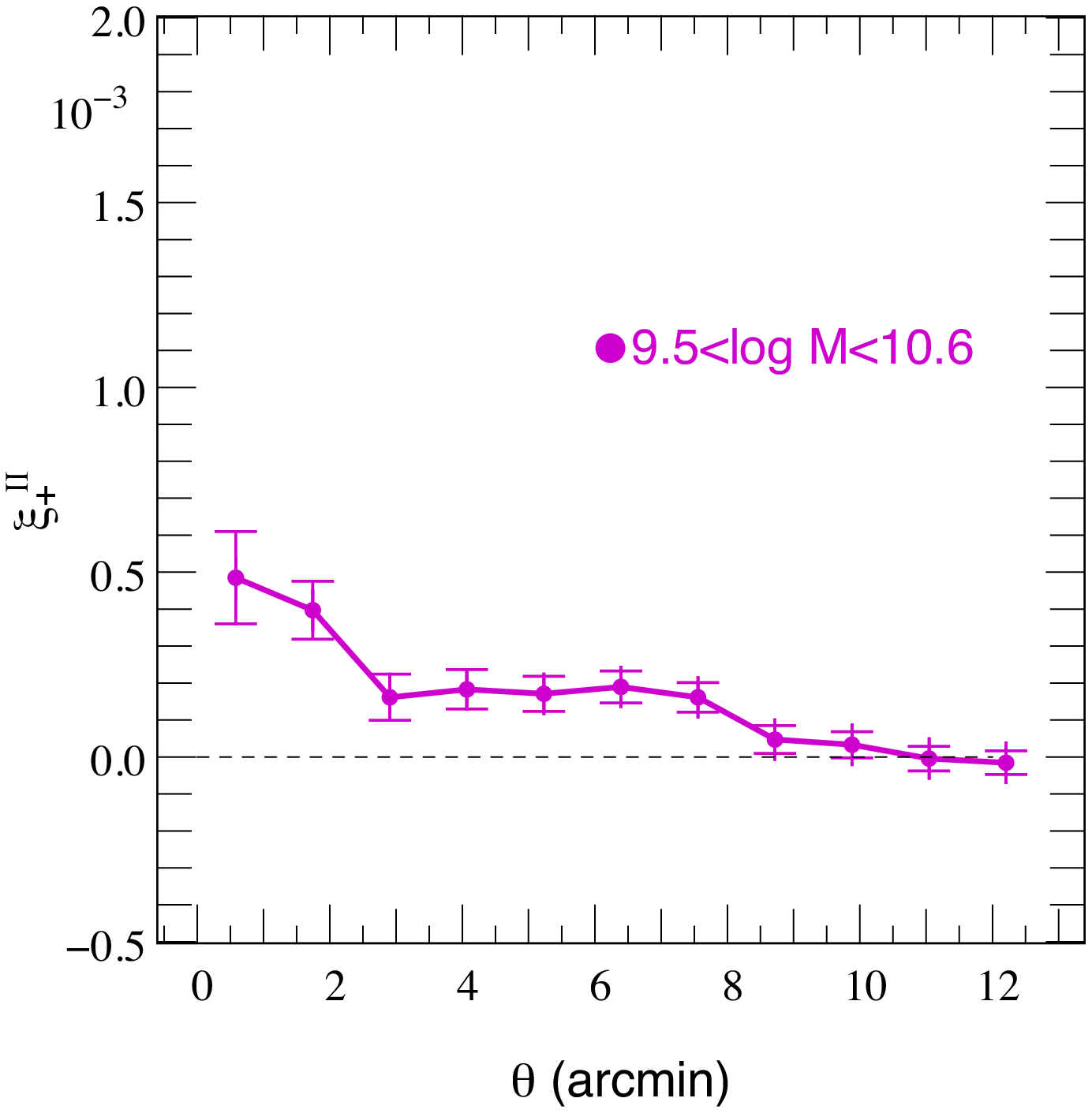}
\includegraphics[width=1.7in]{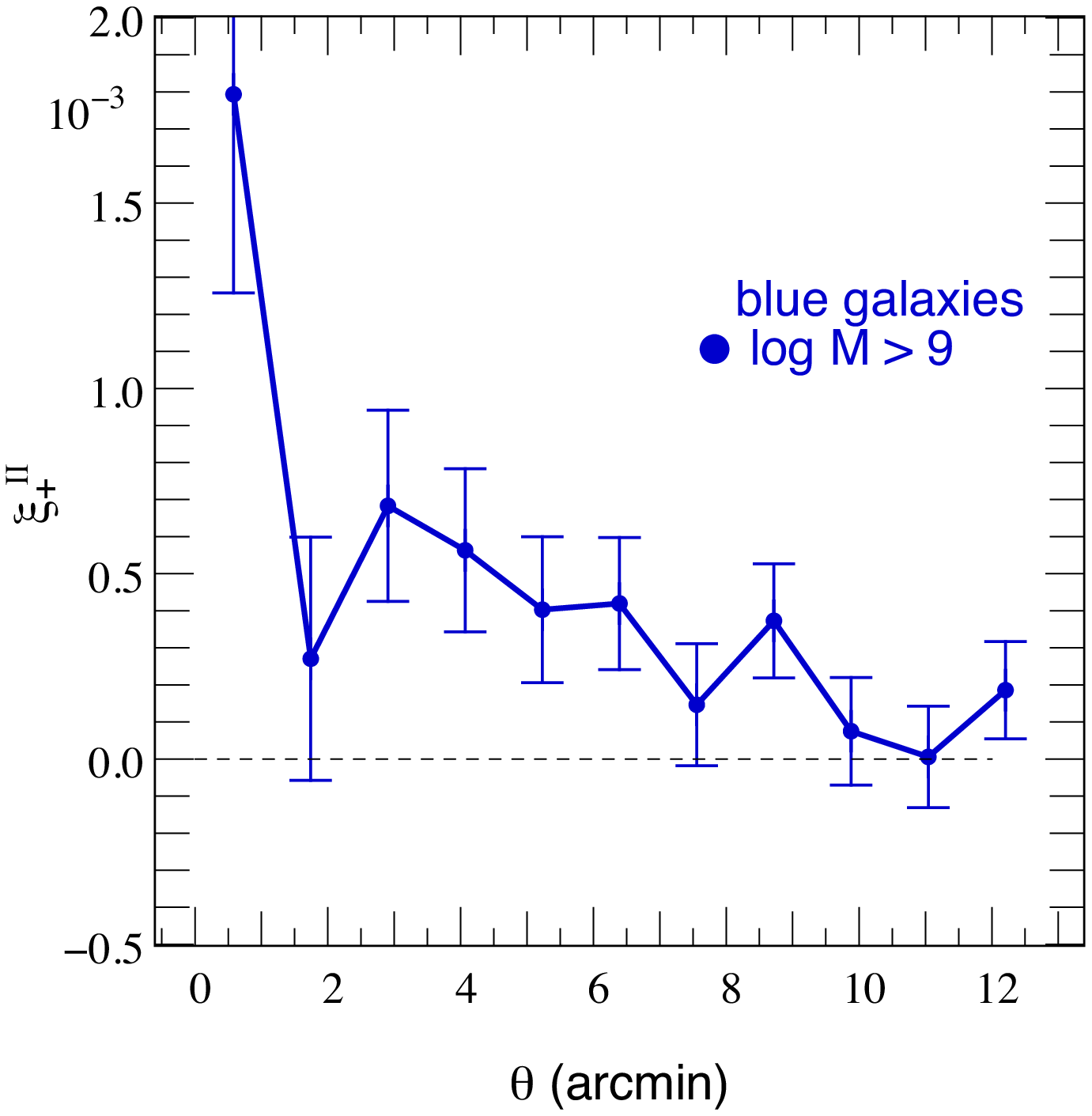}
\includegraphics[width=1.7in]{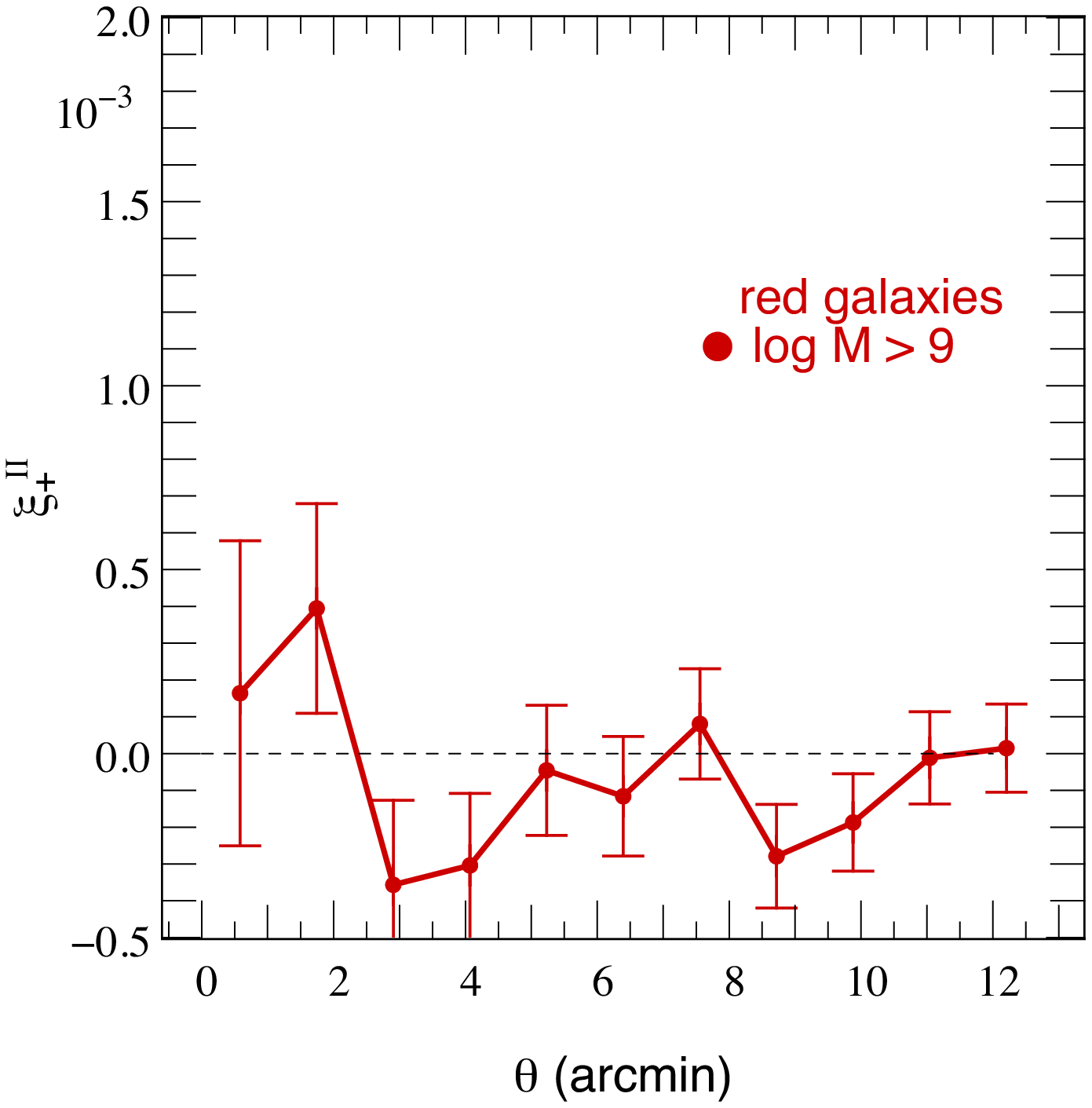}
\caption{
Two-point correlation function of the projected ellipticities of galaxies as a function of angular separation for intermediate mass (left), blue (middle) and red (right) galaxies.}
   \label{fig5}
\end{center}
\end{figure}
In order to get closer to weak lensing observables, one question arises: what fraction of the spin-spin correlations remains after projection on the sky? To address this issue, the spins are projected along a given line-of-sight direction in the box and the apparent axis ratio is assumed to be well-approximated by $q= {\vert L_z\vert}/{\vert \mathbf{L}\vert}$, where $z$ is the line of sight direction.
 The orientation of the major axis of the projected ellipse is $\psi=\pi/2-\arctan (L_y/L_x)$ so that the complex ellipticity can be written $e=(1-q)/(1+q)\exp (2 i \psi)$ in cartesian coordinates. The projected ellipticities can easily be mapped from cartesian $(x,y)$ coordinates to the $(+,\times)$ frame attached to the separation of a given galaxy pair according to the geometric transformation
$
e_+ = -e_x \cos(2\beta) - e_y \sin(2\beta)\,$, 
$e_\times =e_x \sin(2\beta) - e_y \cos(2\beta),
$
where $\beta$ is the angle between the separation and the first cartesian coordinate ($x$).
With those prescriptions, we can estimate the projected correlation functions for a given projected separation $\theta$. For the II component (dropping the subscript ${s}$), this reads
\begin{equation}
  \xi_{+}^{\rm II}(\theta)=\left\langle e_{+}e'_{+}+e_{\times}e'_{\times}\right\rangle\,.
\end{equation}
This correlation function of the projected spins is displayed on Fig.~\ref{fig5} for different samples of galaxies. The spins of blue and intermediate-mass galaxies are shown to be correlated on scales about 10 arcminutes while (as expected from the 3D study) the signal for red galaxies is compatible with zero. Note that
this signal is not contradictory with current observations as it is at a larger redshift.

\section{Conclusion}
In the context of high-precision cosmology (Euclid, DES, LSST, etc), it is crucial to study systematic effects like IA that could significantly contaminate weak lensing observables. \cite{codisetal14} found that at redshift $z=1.2$ in the \hagn 
hydro-dynamical simulation, galaxy ellipticities are correlated with the tidal field and with themselves on cosmological scales with a level of correlation that depends on mass and colour. After projection, these correlation pervades in particular for blue and intermediate-mass galaxies and could be a major source of contamination for cosmic shear studies.

The post-processing of hydrodynamical simulations represent a novel approach to deal with IA which, unlike semi-analytical modeling or linear theory, takes into account  baryonic physics. Mass, colour-dependence and any other selection effects can be modeled  accordingly.
The analysis presented here (see also \cite[Codis et al. 2014]{codisetal14}) is a first step in the accurate modeling of IA effects and paves the way to future more realistic studies (light-weighted measurements on the light cone, etc.).
 
 \vskip 0.5cm
{\sl  This work is partially supported by grant ANR-13-BS05-0005 of the french ANR.
SC thanks Raphael Gavazzi and Karim Benabed for fruitful comments.}

\end{document}